\documentclass[twocolumn,twoside,slac_two]{revtex4}
\usepackage{graphicx}
\usepackage{fancyhdr}
\usepackage{xcolor}
\def \jp {J/\psi}
\def \chicj {\chi_{cJ}}

\newcommand{\ks}{K_S^0}

\newcommand{\BR}{\mathcal{B}}

\newcommand{\ppbar}{p \overline{p}}

\newcommand{\g}{\gamma}

\newcommand{\psp}{\psi(2S)}
\newcommand{\psip}{\psp}

\newcommand{\jpsi}{J/\psi}

\newcommand{\chicz}{\chi_{c0}}
\newcommand{\chico}{\chi_{c1}}
\newcommand{\chict}{\chi_{c2}}
\newcommand{\chicJ}{\chi_{cJ}}

\newcommand{\pipi}{\pi^+\pi^-}
\newcommand{\kskp}{K^0_S K^+ \pi^- + c.c.}

\newcommand{\piz}{\pi^0}

\newcommand{\gpppr}{\gamma \pi^+\pi^-p\bar{p}}
\newcommand{\gkkkk}{\gamma K^+K^-K^+K^-}

\newcommand{\fptk}{2(\pp)\kk}
\newcommand{\kk}{K^+K^-}
\def \pp {\pi^+\pi^-}
\def \kk {K^+K^-}

\def\Journal#1&#2&#3(#4){#1{\bf #2}, #3 (#4)}

\def\PRD{{\it Phys.  Rev.}  { D }}

\def\etal{{\it et al.}}

\begin{document}

\title{Experimental charmonium decay results from BES}

%

\author{Rong-Gang Ping$^1$, F. A. Harris$^2$}
\affiliation{$^1$Institute of High Energy Physics, Chinese Academy
of Science,
Beijing 100049, People's Republic of China\\
$^2$Department of Physics and Astronomy, University of Hawaii, Honolulu, Hawaii 96822, USA}

\begin{abstract}
Based on 14 million $\psip$ and 58 million $\jp$ events collected by
the BESII detector, the leptonic decay of $\psip$ into $\tau^+\tau^-$,
$\psip$ multi-body decays, $\chicj$ decays, and $\jp$ hadronic decays
are studied, and the branching fractions of these decays are
reported. These results may shed light on the understanding of QCD.
\end{abstract}

\maketitle

\thispagestyle{fancy}


\section{Introduction}
The Beijing spectrometer (BES) is a general purpose solenoidal
detector at the Beijing Electron Positron Collider (BEPC) storage
ring, which operates at center-of-mass energies from 2 to 5 GeV. The
BES detector (BESII) is described in Ref.~\cite{detector2}.

In this paper, we focus on studies of the $\psip$ leptonic decay,
$\psip$ radiative decays, $\psip$ hadronic decays, $\chicj$ decays,
and $\jp$ decays based on 14 million $\psi(2S)$ and 58 million $J/\psi$
events collected by the BESII detector.
\section{\boldmath $\psip$ decays}
\subsection{\boldmath $\psi(2S) \to \tau \tau$}
 The $\psi(2S)$ provides an opportunity to compare the three lepton
generations by studying the leptonic decays $\psi(2S)\to e^+e^-,
~\mu^+\mu^-$, and $\tau^+\tau^-$. The leptonic decay widths are
approximately described by the relation $B_{ee}\simeq B_{\mu\mu}\simeq
B_{\tau\tau}/0.3885$, which is in good agreement with the BESI
$B_{\tau\tau}$ measurement~\cite{besitt}.  Based on 14 million $\psi(2S)$
events, the branching fraction for $\psi(2S)\to\tau^+\tau^-$ is
remeasured~\cite{bestaotao}. The $\tau^+ \tau^-$ pairs are
reconstructed from $\tau^+\tau^- \to\mu^+\bar \nu_\tau\nu_\mu e^-
\nu_\tau\bar\nu_e$ and $\tau^+\tau^-\to e^+\bar{ \nu}_\tau\nu_e \mu^-
\nu_\tau\bar\nu_\mu$. At the $\psi(2S)$ resonance, 1015 signal events are
observed, while 516 events are estimated to be from $e^+ + e^- \to \tau^+
\tau^-$,  determined using a
data sample taken at $\sqrt s=3.65$~GeV. The branching fraction is
calculated to be $(0.310\pm0.021\pm0.038)\%$, where the first error is
statistical and the second is systematic. Compared with the BESI
result, the number of events is much bigger and the QED
contribution and the efficiency and background estimations are improved.

\subsection{Radiative decays}
Besides the conventional meson and baryon states, QCD also predicts
a rich spectrum of glueballs, meson hybrids, and multi-quark states
in the 1.0 to 2.5~$\hbox{GeV}/c^2$ mass region. Therefore, searches
for evidence of these exotic states play an important role to
test QCD. Such studies of these states have been performed in
$\jpsi$ radiative decays for a long time~\cite{Jdecay, QWG}, while
studies in $\psip$ radiative decays have been limited due to low
statistics in previous experiments~\cite{PDG, QWG}. The radiative
decays of $\psip$ to hadrons are expected to contribute about 1\% to
the total $\psip$ decay width~\cite{PRD-wangp}. However, the
measured channels only sum up to about 0.05\%~\cite{PDG}.

\begin{figure}\centering
\includegraphics[width=0.45\textwidth, height=11.0cm]{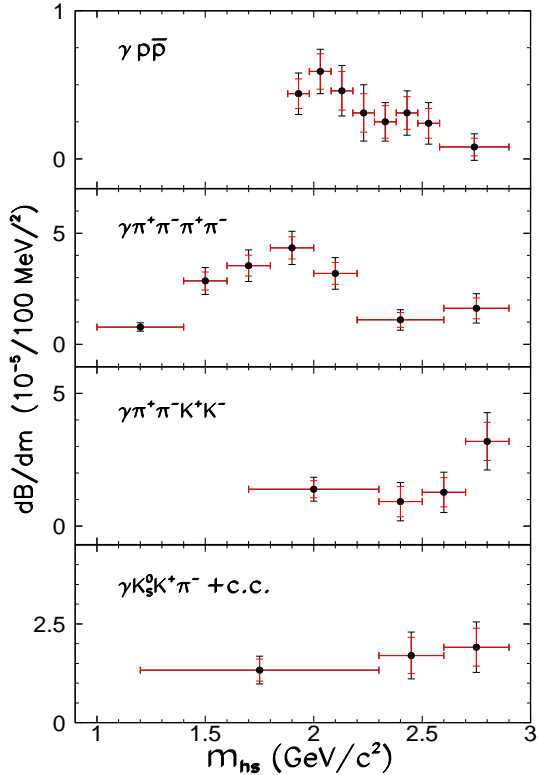}
\caption{ \label{difbr} Differential branching fractions for $\psip$
decays into $\gamma\ppbar$, $\gamma 2(\pipi)$, $\gamma K^+ K^-
\pipi$, and $\gamma \kskp$. Here $m_{hs}$ is the invariant mass of
the hadrons in each final state. For each point, the smaller
longitudinal error is the statistical error, while the bigger one is
the sum of statistical and systematic errors. }
\end{figure}

Recently BESII measured  the decays of $\psip$ into
 $\gamma\ppbar$, $\gamma 2(\pipi)$,
$\gamma \kskp$, $\gamma K^+ K^- \pipi$, $\gamma K^{*0}K^-\pi^+
+c.c.$, $\gamma K^{*0}\bar K^{*0}$, $\gamma\pipi\ppbar$, $\g2(\kk)$,
$\gamma3(\pp)$, and $\gamma 2(\pi^+\pi^-)K^+K^-$,  with the
invariant mass of the hadrons ($m_{hs}$) less than 2.9
$\hbox{GeV}/c^2$ for each decay mode \cite{bes2rad}. The
differential branching fractions are shown in Fig.~\ref{difbr}. In
the mass distribution of $m_{\ppbar}$ for $\psip\to\gamma\ppbar$,
although there is some excess of events near $\ppbar$ threshold as
shown in Fig.~\ref{difbr}(a), no significant narrow structure due to
the $X(1859)$, observed in $J/\psi \to \gamma p \bar{p}$, is seen.
The upper limit on the branching fraction is measured to be
$\mathcal{B}[\psip\to\gamma X(1859)\to\gamma\ppbar]<5.4\times
10^{-6}$ at the 90\% C.L. There is a wide peak in the
$m_{2(\pi^+\pi^-)}$ distribution located at $1.4\sim 2.2$
$\hbox{GeV}/c^2$, but its structure can not be resolved due to the
low statistics. No obvious structure is observed in the other final
states. The branching fractions below $m_{hs}<2.9$
$\textrm{GeV}/c^2$ are given in Table \ref{Tot-nev}, and they sum up
to $0.26\%$ of the total $\psip$ decay width. This indicates that a
larger data sample is needed to search for more decay modes and to
resolve the substructure of $\psip$ radiative decays.

\begin{table}
\begin{center}
\caption{\label{Tot-nev} Branching fractions for $\psip\to\gamma
+hadrons$ with $m_{hs}<2.9$ $\hbox{GeV}/c^2$, where the upper limits
are determined at the 90\% C.L. \cite{bes2rad}.}

\begin{tabular}{ll} \hline \hline
Mode & $\BR(\times 10^{-5})$\\\hline
$\gamma p\bar{p}$ & 2.9$\pm$0.4$\pm$0.4 \\
$\gamma 2(\pi^+\pi^-)$ & 39.6$\pm$2.8$\pm$5.0\\
$\gamma K^0_S K^+\pi^-+c.c.$  & 25.6$\pm$3.6$\pm$3.6 \\
$\gamma K^+ K^-\pi^+\pi^-$ & 19.1$\pm$2.7$\pm$4.3 \\
$\gamma K^{*0} K^+\pi^-+c.c.$& 37.0$\pm$6.1$\pm$7.2\\
$\gamma K^{*0}\bar K^{*0}$&$24.0\pm 4.5\pm 5.0$\\
$\gpppr$& 2.8$\pm$1.2$\pm$0.7 \\
$\gkkkk$ &  $<4$\\
$\gamma3(\pp)$&  $<17$\\
$\gamma2(\pi^+\pi^-)K^+K^-$& $<22$ \\
\hline \hline
\end {tabular}
\end{center}
\end{table}

\subsection{Hadronic decays}
In perturbative QCD (pQCD), hadronic decays of both
$\psi(2S)$ and $J/\psi$ proceed dominantly via the annihilation of
$c\bar c$ quarks into three gluons or one photon, followed by a
hadronization process. This yields the so-called pQCD "12\% rule",
{\it i.e.}
$$Q_h\equiv {B_{\psi'\to h}\over B_{J/\psi\to h}}={B_{\psi'\to
ee}\over B_{J/\psi\to ee}}\simeq 12\%.$$ A large violation of this
rule was firstly observed in decays to $\rho\pi$ and $K^*K+c.c.$ by
Mark II \cite{rhopi}, known as {\it the $\rho\pi$ puzzle}. Since
then there have been many more measurements of $\psi(2S)$ decays by
BES and recently by the CLEO
collaboration for the
study of the 12\% rule. Table \ref{hadbr} summarizes recent measurements on
$\psip$ decays by BES. For the $\psip$ decays listed,
the $Q_h$ ratios are in agreement with the $12\%$ rule within
$1\sim 2\sigma$, except for the obviously suppressed channel
$\psip\to\omega f_2(1270)$.

The branching fractions of $\psip$ decays into octet baryons have also
been measured; the results are listed in Table \ref{hadbr}. They are in
agreement with the results published by the CLEO collaboration within
$2\sigma $ for $\ppbar$ and within $1\sigma$ for the
$\Lambda\bar\Lambda$, $\Sigma^0\bar\Sigma^0$, and $\Xi^-\Xi^+$
channels. For $\psip\to N\bar N\pi$, the ratios of the measured
branching fractions of the $\ppbar \pi^0$ isospin partners  are given by
$\mathcal{B}(\psip\to\ppbar\pi^0):\mathcal{B}(\psip\to p\bar
n\pi^-):\mathcal{B}(\psip\to\bar
pp\pi^+)=1:1.86\pm0.27:1.91\pm0.27$, which is consistent with the
isospin symmetry prediction $1:2:2$.

\begin{table}
\begin{center}
\caption{\label{hadbr} Branching fractions for $\psip$ hadronic
decays. Here the ratio $Q_h$ is defined as $Q_h={\BR(\psip\to
h)\over \BR(J/\psi\to h)}$, where the $\BR(J/\psi\to h)$ values
are
taken from Ref.~\cite{PDG}.}
\begin{tabular}{cccccc} \hline \hline
Mode: $h$  &  $\BR(\times 10^{-4})$ &$Q_h$(\%)\\
\hline
$\ppbar$&$3.36\pm0.09\pm0.25$&$14.9\pm1.4$\\
$\Lambda\bar\Lambda$&$3.39\pm0.20\pm0.32$&$16.7\pm2.1$\\
$\Sigma^0\bar\Sigma^0$&$2.35\pm0.36\pm0.32$&$16.8\pm3.6$\\
$\Xi^-\Xi^+$&$3.03\pm0.40\pm0.32$&$16.8\pm4.7$\\
$p\bar n\pi^-$&$2.45\pm0.11\pm0.21$&$12.0\pm1.5$\\
$\bar pn\pi^+$&$2.52\pm0.12\pm0.22$&$12.9\pm1.7$\\
$ \piz2(\pp)$   &  $24.9\pm0.7\pm3.6$&$10.5\pm2.0$\\
$\omega\pp$     &  $8.4\pm0.5\pm1.2$&$11.7\pm2.4$\\
$\omega f_2(1270)$ & $2.3\pm0.5\pm0.4$ &$5.4\pm0.6$\\
$b_1^\pm\pi^\mp$&  $5.1\pm0.6\pm0.8$ &$17.0\pm4.2$ \\
$\pi^0\fptk$  &  $10.0\pm2.5\pm1.8$  &---\\
 \hline \hline
\end {tabular}
\end{center}
\end{table}

\section{\boldmath $\chicj$ Decays}
\subsection{$\chicj\to\phi\phi$}
 Decays of $\chicj\to
\kk\kk$ are measured using 14 million $\psip$ decays \cite{fk}. The
branching fractions including intermediate states are given in Table
\ref{ppp}. The decay $\chicj\to\phi\kk$ is observed for the first
time, and the precision of the measurements $\chicj\to \phi\phi$ and
$\kk\kk$ are improved compared with PDG values.

The branching fractions of $\chicj\to\phi\phi$ together with
previous BES measurements on $\chicj\to\omega\omega$~\cite{besww} and
$\chicJ\to K^*(892)\bar{K}^*(892)$~\cite{kstarkstar} are used to
predict the decay branching fractions of $\chicJ$ to other vector
meson pairs, like $\rho\rho$ and $\omega\phi$~\cite{zhaoq}, where a large
double OZI suppressed amplitude is expected.

\begin{table*}
\caption {Summary of $\chicj$ hadronic decays. Upper limits are
given at the 90\% C.L. For $\chicj\to\ks K^+\pi^-+c.c.$ and $\eta\pp$,
branching fractions of Br$(\psi'\to\gamma\chicz)=(8.6\pm 0.7)\%$,
Br$(\psi'\to\gamma\chico)=(8.4\pm 0.8)\%$, and
Br$(\psi'\to\gamma\chict)=(6.4\pm 0.6)\%$ are used in the
calculation. For other decays, branching fractions of
Br$(\psip\to\gamma \chicj)$ from CLEOc \cite{cleochicj} are used.} {\begin{tabular}{@{}c|c|c|c@{}}\hline\hline
 Decay mode X& Br$(\chicz\to X)~(\times 10^{-3})$   & Br$(\chico\to X)~(\times 10^{-3})$ &Br$(\chict\to X)~(\times 10^{-3})$ \\\hline
$2(\kk)$&$3.48\pm0.23\pm 0.47$&$0.70\pm0.13\pm 0.10$&$2.17\pm0.20\pm 0.31$\\
$\phi\kk$&$1.03\pm0.22\pm 0.15$&$0.46\pm0.16\pm
0.06$&$1.67\pm0.26\pm
0.24$\\
$\phi\phi$&$0.94\pm0.21\pm0.13$&---&$1.70\pm0.30\pm0.25$\\
 $\ks K^+\pi^-+c.c.$&$<0.35$&$4.0\pm0.3\pm
0.5$&$0.8\pm0.3\pm0.1$\\
 $\eta\pp$&$<1.1$&$5.9\pm0.7\pm 0.8$&$<1.7$\\
\hline\hline
\end {tabular}\label{ppp}}
\end {table*}

\subsection{$\chicj\to \ks K\pi,\eta\pi\pi$}
 Decays of $\chicz$ and $\chict$ into three
pseudo-scalars are highly suppressed by the spin-parity selection
rule. BES measured the branching fractions of
$\chico$ decays into $\kskp$ and $\eta\pipi$, including the intermediate states
involved \cite{mppbar4}. The branching fractions or upper limits at
the 90\% C.L. are summarized in Table \ref{ppp}.

The $\kskp$ events are mainly produced via $K^*(892)$ intermediate
states, and the $\eta\pipi$ events via $f_2(1270)\eta$ or
$a_0(980)\pi$. The branching fractions \cite{mppbar4} for these
resonances are

\begin{eqnarray*}
\textrm{Br}(\chico\to K^*(892)^0\bar
K^0+c.c.)~~~~~~~~~~~~~~~~\nonumber\\=(1.1\pm0.4\pm0.2)\times 10^{-3},\nonumber\\
\textrm{Br}(\chico\to K^*(892)^+
K^-+c.c.)~~~~~~~~~~~~~~\nonumber\\=(1.6\pm0.7\pm0.3)\times 10^{-3},\nonumber\\
\textrm{Br}(\chico\to
f_2(1270)\eta)~~~~~~~~~~~~~~~~~~~~~~~~~~~\nonumber\\=(3.0\pm0.7\pm0.5)\times
10^{-3},\nonumber\\
 \textrm{Br}(\chico\to
a_0(980)^+\pi^-+c.c.\to
\eta\pp)~~\nonumber\\=(2.0\pm0.5\pm0.5)\times 10^{-3}.\nonumber
\end{eqnarray*}

 Except for $\chico\to \kskp$, all other modes are first
observations.
\section{\boldmath $\jp\to\Lambda\bar\Lambda\pi^0,~\Lambda\bar\Lambda\eta$}
The isospin violating process $\jp\to\Lambda\bar\Lambda \pi^0$ has
been studied by the DM2 \cite{dm2} and BESI \cite{bes1}
collaborations, and its average branching fraction is quoted as
$\mathcal{B}(\jp\to\Lambda\bar\Lambda\pi^0)=(2.2\pm0.6)\times 10^{-4}$
\cite{PDG}. However, the isospin conserving process
$\jp\to\Lambda\bar\Lambda\eta$ has not been reported. Is it true that
$\mathcal{B}(\jp\to\Lambda\bar\Lambda\pi^0)>\mathcal{B}(\jp\to\Lambda\bar\Lambda\eta)$?
BESII used 58 million $\jp$ events to study
$\jp\to\Lambda\bar\Lambda\pi^0$ \cite{bes2llp}. It is found that this
decay is seriously contaminated by
$\jp\to\Sigma^0\pi^0\bar\Lambda+c.c.$  and
$\Sigma^+\pi^-\bar\Lambda+c.c.$ decays. For estimating these backgrounds, the
branching fraction of $J/\psi \to \Sigma^+\pi^-\bar\Lambda+c.c.$ was
measured and determined to be to be $(15.2\pm0.8\pm1.6)\times 10^{-4}$.
Assuming isospin symmetry between this and the neutral decay
mode and after
subtracting these backgrounds, no significant signal is observed for
$\jp\to\Lambda\bar\Lambda\pi^0$, and the upper limit is determined to
be $\mathcal{B}(\jp\to\Lambda\bar\Lambda\pi^0)<0.64\times 10^{-4}$ at
the 90\% C.L. A clear signal is seen for
$\jp\to\Lambda\bar\Lambda\eta$ and the branching fraction is determined to be
$(2.62\pm0.60\pm0.44)\times 10^{-4}$. This indicates that the
$\jp\to\Lambda\bar\Lambda\pi^0$ is suppressed due to isospin
conservation violation.  \\
\section{Summary}
Using the 14 million $\psip$ and $58$ million $\jp$ events taken with BESII
detector at the BEPC storage ring, BES Collaboration has provided
many interesting results on charmonium decays, including the
observation of many $\psip$ radiative decays, some $\psip$ hadronic
decays, $\chicj$ decays, and determined that  the isospin
violating process $\jp\to\Lambda\bar\Lambda\pi^0$ is suppressed.

{\vspace{0.5cm}\Large\bf Acknowledgments\vspace{0.5cm}}\\
I thank my colleagues in BES Collaboration for many helpful
discussions. This work was supported in part by National Natural
Science Foundation of China under Contract No. 10491300, and the
Department of Energy under Contract No. DE-FG02-04ER41291
(University of Hawaii).


\end{document}